# The role of attitudinal factors in mathematical on-line assessments:

# a study of undergraduate STEM students

Elizabeth Acosta-Gonzaga*# and Niels R. Walet*

*School of Physics and Astronomy, The University of Manchester, Manchester, M13 9PL, UK

#Instituto Politécnico Nacional (IPN), Av. Té 950, Granjas México, Iztacalco, 08500, Distrito Federal, México

## Abstract

This study explores student attitudes to the use of substantive on-line assessments that require mathematical answers. Our goal is to learn what are the important aspects in a design of more effective e-assessments that support learning of mathematical subjects in a higher education setting. To that end we analyse the effects of a variety of attitudinal factors towards such assessments amongst a cross-section of 1$^{st}$ year students in an English University. These students were all previously exposed to on-line assessments containing substantial mathematical work, including testing of and feedback on the algebraic structure of their answers. They were provided with detailed online feedback, and we therefore specifically examine the effect of formative feedback on the usage of educational technology. Our results suggest that students find on-line feedback more enjoyable and useful than traditional feedback. 'Attitude' and 'Enjoyment' are the two most important factors influencing their usage intention. Our results also show that, even for this digital generation, confidence in using computers and the availability of support for using information technology are important factors in making effective use of on-line assessments.

## 1  Introduction

The continuing expansion of the use of Information and Communication Technology (ICT) gives us many new opportunities, but also poses serious challenges. In the context of learning, a drive towards cost saving and the lure of the new have often led to implementations which are not wholly successful. Many of these implementations have ignored the important attitudinal factors that can affect the usage of learning technology. Information system researchers have identified the importance of such personal factors,e.g., attitudes, beliefs, culture and behaviour, in technology acceptance  ago (Davis, Bagozzi & Warshaw,  1992). Such factors have also been investigated in a general setting for the acceptance of educational technology in higher education (Liu, Liao & Pratt, 2009), (Teo, 2009), (Terzis & Economides, 2011), (Cheung & Vogel, 2013). Little work has been done that applies to those areas that require a rather specialised style of questions, such as in mathematics.

Most on-line material provides a mixture of what we can roughly call "learning materials" and "assessments". The latter contains both formative and summative assessments, sometimes separate, and sometimes intermingled. There are serious difficulties with providing meaningful online assessment for questions requiring mathematics in their solution and/or answer. As a result there is no reason to believe that the factors playing a role in adopting e-learning in a mathematical context have been fully explored by the more generic studies quoted above. This is especially true in an a highly specialised setting such as the use of mathematics in higher education, where we would typically like to assess deeper levels of (mis)understanding. The interesting studies quoted above usually look at simpler types of assessment such as multiple choice questions, and those assessments are not necessarily applicable when we support and analyse the solution and understanding of more complex mathematical problems.

Formative electronic assessment provides students with a flexible opportunity to evaluate their understanding of a course, and a mechanism to set their own learning goals and assess their weaknesses and strengths in order to improve their performance in a course (Wilson, Boyd, Chen & Jamal, 2011). Understanding their strengths and weaknesses enables students to focus on their cognitive development. It will also help them to identify key areas of focus for further study. The work (Miller, 2009) shows that a formative e-assessment framework has the potential to provide an effective mechanism for providing this feedback. that students who use computer-assisted practice quizzes, gain significantly higher grades than students who do not. This demonstrates that computer-assisted formative assessments, and specifically online practice tests, can have a positive impact on student performance. "Effective feedback on formative electronic assessments requires that feedback to students be provided immediately after the student has responded to an item"(Miller, 2009). Feedback it not only improves student learning but also encourages independent learning. Particularly in higher education, formative e-assessments are able to promote self-regulated learning (Gipps, 2005).

Using the computer to assess mathematical subjects can bring important advantages; for instance, (Angus & Watson, 2009) illustrate that when students studying mathematics are subject to regular challenging on-line testing, their learning, as measured by an end-of-semester examination, is significantly improved. Indeed, mathematics is particularly suited to an on-line assessment strategy and on-line assessment can provide valuable feedback to students, particularly distance learners (Whitelock & Raw, 2003), but there are difficulties extending this to the depth required for university students. The benefits of online assessment can obviously be extended to students in blended educational contexts. For example, at a primary school level, (Peltenburg, van den Heuvel-Panhuizen & Doig, 2009) show that e-assessment using a dynamic visual tool is able to support students to overcome difficulties in solving subtraction problems. This type of testing allows the teacher to better examine the students' actions and thinking processes than is possible with a paper-and-pencil test (Whitelock, 2009).

Other researchers such as (Dettori, Garuti & Lemut, 2002) have investigated the possibilities of improving mathematical teaching and learning by using technology in various educational contexts. They report that studying mathematical topics supported by technology leads to a considerable increase in student motivation towards learning mathematics, which also creates a general positive change in their attitude towards the subject (Ursini, Sánchez, & Orendain, 2004). There are several studies that study this attitude (Reed, Drijvers & Kirschner, 2010), (Ursini *et al.*, 2004), (Galbraith & Haines, 1998), which is associated with determining their willingness to learn. In this work we do not consider this attitude per se. Instead, the focus is on analysing the attitude and intention of *using* electronic assessment technologies to learn mathematics in the specific context of higher education.

Our approach is rooted in the belief that on-line assessment can be a positive influence on the learning process, provided students have the right attitude towards using computers for learning. Generating the appropriate attitude towards using computers can not only play an influential role in determining the extent to which students accept it as a learning tool, but can also determine future behaviour such as using a computer for further study and vocational purposes (Rosen & Weil, 1995). This coincides with what is mentioned in (Venkatesh, Morris, Davis & Davis, 2003), where it is suggested that users' attitudes are a key factor in predicting technology acceptance for future use. Therefore, it is important to understand what drives and stimulates students to use on-line testing in mathematical subjects.

We also examine the role of formative on-line feedback on the usage of educational technology. As we said before, the important role of assessment and feedback in learning and teaching in higher education has been well recognised in literature (Brown, Bull & Pendlebury, 1997), (Gibbs & Simpson, 2004), (Nicol & Macfarlane-Dick, 2006), (Bloxham & Boyd, 2007). It is in this process that students' learning gets consolidated, which in turn produces persistent changes in students understanding. This is the reason why educational assessment deserves such attention. The advantages of on-line assessment technologies, either formative or summative, have been summarised by, e.g., (Terzis, Moridis & Economides, 2012): "(A) high interaction and adaptation with test-takers, (B) real-time feedback, (C) real-time score reports, (D) more efficient management, setting, and delivering of exams, (E) easier data management, (F) cost reduction, (G) self-evaluation and recognition of students' strengths and weaknesses". Also, in order to provide good quality learning, universities can make use of innovative technologies to support the teaching, learning and assessment processes. They should clearly ask what should the role of information technologies be? One of the drivers is providing access and support to more and more students. Given that undergraduate classes can consist of several hundred students, it is no longer possible for the faculty to meet with individual students and guide their learning. This issue combined with the diversity of students' academic backgrounds supports the need for formative electronic assessment (Miller, 2009). Immediate on-line feedback is an essential feature of such assessments: It has positive effects on students' learning performance; it activates their intrinsic motivation (Dreher, Reiners & Dreher, 2011) and helps them to achieve their goals (Whitelock & Brasher, 2006). Since effective use of on-line feedback is clearly desirable for all stakeholders, it is important to explore the students' perception of the on-line feedback. In this study we specifically explore the effects of on-line feedback on 'usefulness' and 'enjoyment'.

In this paper we report the results of a study carried out with first year undergraduate students at the University of Manchester, UK, who are studying STEM subjects. We studied the attitudinal factors using an online questionnaire, structured according to a set of empirical and theoretical factors discussed in Section 2. We construct a model from these factors in Section 3. We then discuss the data collection in Section 4, and then analyse the data using our models in the next section. Finally, we draw some conclusions and give some recommendations. In the first appendix we show our full questionnaire, and in the second we illustrate some of the essential features of the type assessments we are discussing.

## 2 Theoretical background

The model developed and applied in this work is empirical, but we use prior studies as a guiding principle in designing it. We wish to understand the influences on a student's 'behavioural intention'. We start the development of our model from the Technology Acceptance Model (TAM), first proposed by (Davis, 1989) as a model to predict information systems adoption. TAM itself is based on the earlier Theory of Reasoned Action (TRA). TRA was first proposed by (Fishbein & Ajzen, 1975) as a model to explain and predict the behaviour of people in a specific situation. TRA states that a person's actual behaviour can be explained by his/her intention and beliefs, and that, intentions can be explained by both his/her 'attitude' and 'subjective norms'. In TRA 'attitude' is defined as "the degree of a person's favourable or unfavourable evaluation or appraisal of the behaviour in question". (Fishbein & Ajzen, 1975) define the 'subjective norms' as the "person's perception that most people who are important to him or her think he or she should or not should perform the behaviour in question". The effects of 'subjective norms' on behavioural intention are direct. The TRA also states that 'attitude' plays an important role in a person's 'behavioural intention'. These two factors were later adopted by the TAM.

In the TAM a few new factors that may have an effect on adoption are defined. The most important ones are 'perceived usefulness' (PUS) which is defined as "the degree to which an individual believes that using a particular system would enhance his/her job performance", and 'perceived ease of use' (PEU), which is defined as "the degree to which an individual believes that using a particular system would be free of physical and mental effort" (Davis, 1989). The TAM also assumes that 'perceived usefulness' itself will be influenced by 'perceived ease of use', because if "two systems perform the identical set of functions, a user should find the one that is easier to use more useful" (Davis, 1993). TAM states that 'perceived usefulness' has a direct effect on an individual's behavioural intention toward using a system, and 'perceived ease of use' acts indirectly through its influence on 'perceived usefulness' (Davis, 1989). That means that 'perceived usefulness' mediates the effect of 'perceived ease of use' on 'behavioural intention'. It is important to consider these factors in our work, since it is known that both 'perceived usefulness' and 'perceived ease of use' can have significant effects on an individual's behavioural intention to use e-learning systems (Liu *et al.*, 2009); (Ong, Lai, & Wang, 2004).

Borrowing from social cognitive theory (SCT) (Bandura, 1986), we also include the construct of 'self-efficacy' (SE) in our model. The SCT is a widely accepted and empirically validated theory for understanding and predicting human behaviour and for identifying methods by which behaviour can be changed. It has been applied as a theoretical framework to predict and explain an individual's behaviour in a variety of contexts involving cognitive, social, motor, health, instructional, and self-regulatory skills. The SCT proposes two cognitive factors: 'self-efficacy' and performance expectations that influence individual behaviour (Bandura, 1986). Performance expectations are defined as "the degree to which a learner believes that using BELS will help him or her to attain gains in learning performance". The definition is similar to the concepts of perceived usefulness, based on Davis's technology acceptance model, which we have already included. In this context, 'self-efficacy' is defined "as one's judgements and beliefs of his/her confidence and capability to perform a specific behaviour" (Bandura, 1986). 'Self-efficacy' has been shown to be able to predict behavioural change with different types of participants in various settings. Therefore, in order to understand its impact we include it as a factor in our model. In the context of our study it is included as 'computer self-efficacy' (CSE).

It is well-known that external factors can have an important influence on attitudes; in this context think about, e.g., the accessibility of IT services. We have therefore included the 'availability of IT' (AIT) as a potential factor. It is similar to what (Thompson, Higgins, & Howell, 1991) and (Venkatesh *et al.*, 2003) call "technology and resource facilitating conditions" and (Ajzen, 1991) mention as "perceived behavioural control". In the context of PC use, (Thompson *et al.*, 1991) define facilitating conditions as "the provision of support for users may be one type of facilitating condition that can influence system utilization. By training users and assisting them when they encounter difficulties, some of the potential barriers to use are reduced or eliminated". This is consistent with what (Venkatesh, 2000) explains "in the context of workplace technology use, specific issues such as the availability of support staff, which is an organizational response to help users overcome barriers and hurdles to technology use, especially in the early stages of learning and use". In other words, the facilitating conditions include the factors in the environment that shape a person's perception of ease or difficulty of performing a task (Teo, 2012). It embraces factors such as technical support (the provision of help-desks and on-line support services). Technical support has been cited as one of the important factors in the acceptance of technology for teaching and in user satisfaction (Williams, 2002), (Teo, 2012).

By analysing the process of e-assessment, we can identify the key processes, and thus potentially important factors. One such process is the provision of formative feedback, a process where teachers obtain information about the state of a student's learning, and then use this information to determine specific strategies in order to adjust or improve student's learning. The terminology 'formative assessment' is used to indicate the continuous process of communication between the teacher and the student. In this process teachers evaluate students' performance by reviewing student assignments, making corrections and giving suggestions to students in order to improve their assignments. The objective of formative assessment is to make students aware how effective they are as learners through reflection and regular feedback. Of course, we can also use the computer to provide such feedback. Students can learn from their own reflection (self-reflection) in a self-assessment process and/or from peer interactions (peer-assessment) (Hodgson & Pang, 2012). Formative assessment can have a remarkable impact on how students acquire knowledge and build abilities and skills. Through reflection on their own learning, students can assess their activities, by comparing them with the standard required in order to determine if an adjustment is necessary.

In order to examine the role that feedback plays in student attitudes, we analyse two constructs related to the formative assessment process. The first construct captures the students' experience with receiving on-line feedback. We have asked for aspects such as the on-line feedback being precise, clear, helpful and timely. This is combined in a factor called 'received feedback' (RF) which stands for "the student's belief that the support and feedback they receive from the on-line platform will enhance their learning".

(Dreher *et al.*, 2011) remark on the pedagogical benefits obtained, particularly surrounding feedback practices, by using e-assessment technologies. Technological tools allow students can in turn be freed to determine their own learning path along defined milestones and support them by assessing their learning for successful performance. They mention that the real benefit for students is getting immediate feedback, which enhances their learning performance and also stimulates their intrinsic motivation. This matches with (Brasher & Whitelock, 2006)) who mention that formative assessment practices using computers encourage immediate feedback, which has positive effects for students in achieving their goals.

The generic advantages of providing on-line feedback have already been analysed, for instance, by (Terzis, 2012). It is pointed out that the main reasons that learners enjoy using computer-based assessments are: "(1) Learners are able to take the assessment anywhere and any-time using a computer. (2) They are able to take the test as many times as they wish. (3) They feel confident regarding the results' accuracy and fairness since the computer does not care who the test taker is. (4) They are able to see their results as soon as they complete the assessment (Cassady & Gridley, 2005). (5) Electronic assessment provides them immediate feedback that helps identify their strengths and weaknesses" (Wilson *et al.*, 2011), (Crippen & Brooks, 2002). This means that by using feedback teachers can close the distance between a student's current and desired performance (Heinrich, Milne & Moore, 2009).

Therefore, our second construct explores the difference between receiving on-line feedback and face-to-face feedback. In others words, we want to obtain students' opinions regarding the differences of getting each type of feedback. We define 'comparative feedback' (CF) as "the student's perception of receiving on-line feedback when it is compared with traditional feedback". We hypotize that on-line feedback will have a direct positive effect on usefulness and enjoyment (and indirect effects on attitude and behavioural intentions to use web-based testing).

It should thus come as no surprise that we also include an explicit 'enjoyment' construct, since enjoyment is known to be a significant reason for interest in using a specific system. Although research on its impact on behavioural aspects comes from an organizational context, (Davis, 1989), (Davis, 1993) it looks to be generally relevant. In an educational setting, enjoyment has been shown to have a positive influence on usage intentions. (Venkatesh, 2000) explains that playfulness is related to intrinsic motivation or the "perceptions of pleasure and satisfaction from performing the behaviour". This also matches with the work of (Davis *et al.*, 1992) who argue that intrinsic motivation refers to the performance of an activity for no apparent reward other than the process of performing the activity per se. In his research (Venkatesh, 2000) explains that playfulness is a construct that is system-independent, supporting this by the fact that users who are more "playful" with computer technologies in general enjoy using a new system just for the sake of using it. Playfulness is expected to be a relevant factor even when the systems perform a rather boring task, since it still involves exploration and discovery. Venkatesh states that, from a theoretical standpoint, a higher level of playfulness will lead to a lower perception of effort. In the context of this research, this could mean that students feel they have to invest less effort when they perceive an electronic educational task (mathematical exercise) to be enjoyable. If we can consider students as "digital natives", accustomed to using technology in all aspects of their life, it must be relevant to explore the effects of this construct on attitude and intention to use. This could thus be a direct determinant to enhance learning of mathematics.

On the other hand, (Davis *et al.*, 1992) find that 'perceived enjoyment' and 'perceived usefulness' mediate the influence of 'perceived ease of use' on 'intention', explaining that "while usefulness will once again emerge as a major determinant of intentions to use a computer in the workplace, enjoyment will explain significant variance in usage intentions beyond that accounted by usefulness alone." This clearly requires explanation and investigation. In the context of computer-based assessment, the research of (Terzis & Economides, 2011) includes perceived playfulness and shows a positive effect on the behavioural intention to use the system. Therefore, we will explore the effects of this construct on attitude and behavioural intentions.

Table 1: Abbreviated labels of the constructs used in our model, together with their full names

| construct | meaning |
|---|---|
| PUS | Perceived Usefulness |
| PEU | Perceived Ease of Use |
| CSE | Computer Self-Efficacy |
| SI | Social Influence |
| AIT | Availability of IT Services |
| EN | Enjoyment |
| RF | Received Feedback |
| CF | Comparative Feedback |
| AT | Attitude |
| BI | Behavioural Intention |

# 3 Research model and hypotheses

All the constructs proposed in the previous section are included in a hierarchical model, taking into account the network of TAM. The constructs are summarized in Table 1. From these constructs, using the theoretical and empirical approach from the previous section, we then construct a model as displayed in Fig. 1. This model is based on an initial exploration, and contains 14 hypothetical causal relationships between constructs, that will be tested on a set of data using a so-called recursive Structural equation model (SEM). This assumes that each variable has a linear relation with any antecedent variables, and the recursive nature states that causation is unidirectional. Variables are usually divided into tow classes: exogeneous, or explanatory variables $x$ and endogenous or response variables $y$.

We assume a linear form relation between the variables,

$$\sum_{j=1}^{q} B_{ij} y_j + \sum_{j=1}^{p} \gamma_{ij} x_j = \eta_i$$

where $B$ is a lower-triangular matrix,

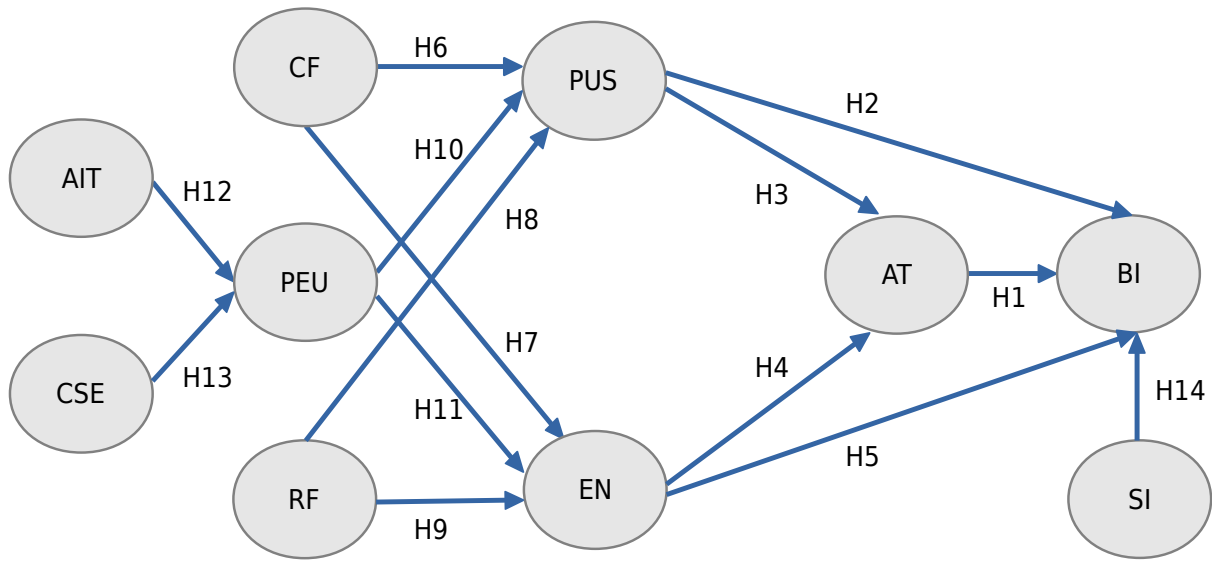

*Figure 1. A visual representation of the model for student responses. The arrows represent hypothetical relationships between the latent variables. For the meaning of the labels, see Table 1.*

$$B_{ij} = \begin{pmatrix} 1 & 0 & 0 & \cdots \\ b_{21} & 1 & 0 & \cdots \\ b_{31} & b_{32} & 1 & \ddots \end{pmatrix}$$

and $\eta$ is a random fluctuation. We then try to explain as much variance in each construct as possible by the antecedent variables, which is what is called the partial-least squares SEM, or PLS-SEM.

# 4 Background of collected data

We invited a large number of first year students (about 1000) in the Faculty of Engineering and Physical Sciences at the University of Manchester to answer an anonymous on-line questionnaire. These students are exposed to a variety of on-line materials. Particularly relevant to the subject of this research is the use of mathematical questions using the Stack engine within a Moodle platform (Sangwin, 2013) that provides feature-rich on-line assessments with malrule based detailed feedback. Specifically to the School of Physics and Astronomy, from which a large fraction of the respondents originate, students also experience online assessment using the "mastering physics" product originally developed by MIT, and now marketed by Pearson. For more details of the way this is used, see (Walet & Birch, 2012).

Every year, the faculty receives a large number of STEM students. As the incoming students have different mathematical backgrounds, we determine their previous level of mathematical knowledge using a diagnostic test. Practice material for the test, and follow-on tests and remedial material, are all provided within Stack. This allows us to differentiate the material delivered to students and provide additional support when required based on preparation and needs.

Students have the option to practise a large number of online mathematical exercises before answering the diagnostic test. The key characteristic of this material is to provide students with immediate feedback through a web-based platform. If a student answers a question incorrectly, they will be given support based on their mistakes by providing a series of hints to help them obtain the correct solution. Students can practise as long as they want and need to reinforce their mathematical skills. This formative feedback represents a big advantage for on-line testing since it opens the possibility of providing students with customised feedback given that the computer can generate this based on the answer given by the student (López, 2009).

While academics may have a wide variety of reasons for selecting particular assessment methods, they need to be aware of their students' perceptions of these methods and how these influence students' learning (Iannone & Simpson, 2013). However, in order to obtain a complete view of the context, it is also important to consider what students' think about learning mathematics through technology. What is their opinion? In order to make the most of what technology offers, students should be willing to undertake mathematical exercises and exams using technology to make their learning of mathematics more valuable.

All students in STEM subjects at the University of Manchester will have taken mathematics up to final high-school exam level (except in the case of Chemistry, where this not required). In the English system, that means a mathematics A-level. Each school has its own requirements for entry, and requires different grades in the exams. Nevertheless, all students are required to take the same diagnostic test, and use the same follow-on material. Most students follow mathematics courses in the School the Mathematics, apart from physics students who follow their courses in the School of Physics and Astronomy. In all cases, a variety of on-line learning materials is used. It is important for the Schools to obtain enough data to build a customised learning strategy to provide students with personalised good quality teaching.

Students used the on-line platforms during September and October 2014. At that point we used an on-line survey to gather students' responses. We obtained 121 full responses. The gender balance is 25.62% (31) female 74.38% (90) male, roughly in line with the gender balance in the faculty. The distribution across fields of study is displayed in Table 2.

*Table 2: Distribution of respondents by School of study*

| School | Responses | Percentage |
|---|---|---|
| Chemical Engineering & Analytical Science | 20 | 16.53% |
| Chemistry | 3 | 2.48% |
| Earth, Atmospheric & Environmental Sciences | 2 | 1.65% |
| Electrical and Electronic Engineering | 14 | 11.57% |
| Mathematics | 6 | 4.96% |
| Mechanical, Aerospace and Civil Engineering | 11 | 9.09% |
| Physics & Astronomy | 65 | 53.72% |

For all 31 questions we invite answers on a five-point likert scale, where in every case the most negative answer is coded as 1, and the most positive as 5. Using this coding, we show the average and standard deviation for each of the questionsIn Fig. 2 (see Table 5 for the details of the questions). As we can see—this will be studied further below—we have very similar results for the individual questions making up a single construct.

# 5 Findings and Results

We use Structural Equation Modelling (SEM) as a statistical technique to analyse relationships between constructs

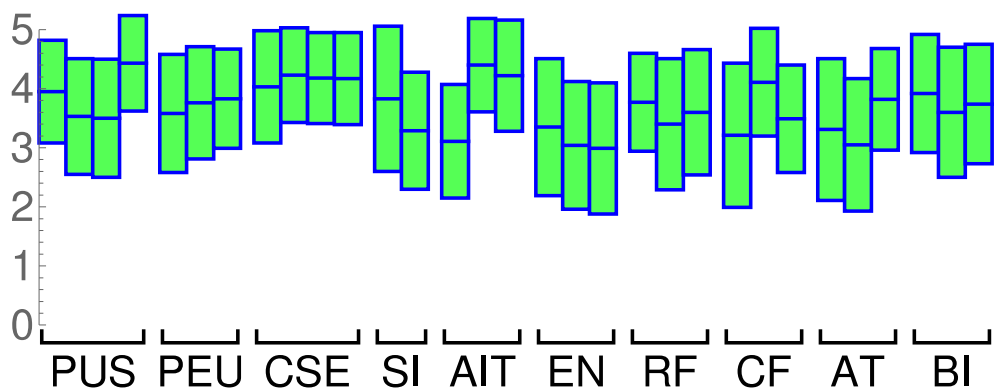

*Figure 2. Descriptive statistics for the questions used for each item: The green bar denotes the standard deviation (+/-) on each item; the line in the middle of the green bars denotes the average*

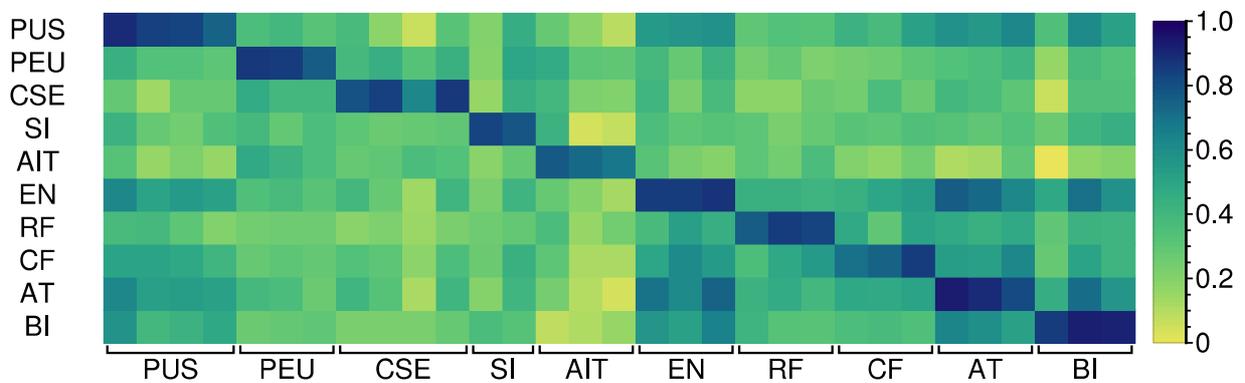

*Figure 3. Indicator loading and cross-loading coefficients as a measures of indicator reliability.*

(hypothetical latent variables). SEM is a widely used multivariate approach. This method can be used to examine the relationship between theory-based latent variables and their indicator variables by using measurements of directly observable indicator variables (Hair, Black, Babin & Anderson, 2010). Applying SEM uses a two-tier process. First the individual measurements or indicators (in our case answers to individual questions) are combined into constructs or latent variables—the consistency of this process needs to be validated for all latent variables. It measures how well the observed indicators fit the unobserved (latent) variables. In the second stage of the process, we analyse a structural model, which covers the relationships among constructs. The constructs typically represent feelings, attitudes, and opinions of a person. The relationships between constructs are hypothesized in accordance with theoretical and logical reasoning (Götz, Liehr-Gobbers & Krafft, 2010). In this work both the measurements and the structural models were evaluated using a PLS-SEM analysis using SmartPLS 2.0 (Ringle, Wende & Will, 2005).

## 5.1 Measurement Model.

We first test the quality of the measurement model. Table 3 shows measures of construct and composite reliabilities. We apply some standard tests that are not normally part of the measurement as well. The values of Cronbach's alpha, which can be used to analyse the correlation between questions making up a construct, show that all constructs have a good internal reliability except "comparative feedback", which with a value α=0.66 can only be considered as acceptable. The "availability of information technology" has a value of alpha that suggests the result is poor. The SEM code evaluates composite reliability and average variance extracted (AVE) as alternative tests of the measurement model. Analysing the values for composite reliability, we find these are all larger than the acceptable cut-off of 0.6 (Bagozzi & Yi, 1988). The Average Variance Extracted ranges from 0.54 to 0.81. These values all exceed the recommended lower cut-off (Götz, Liehr-Gobbers, & Krafft, 2010). Therefore, all values obtained confirm what is called "convergent validity", the internal consistency of each latent variable. Table 3 shows values from the measurement model demonstrating that all indicator loadings are higher than the common threshold criterion of 0.7 to reach indicator reliability. We also show the discriminant validity—the fact that the different indicators are statistically independent.

Table 3: The first three columns show construct and convergent validity coefficients for the measurement model. The remaining columns show the discriminant validity as the off-diagonal entries, which is given by bivariate correlations. On the main diagonal, we show the related square-root of AVE (Average Variance Extracted) of each construct.

|  | Descriptive statistics | | | Bivariate correlations | | | | | | | | | |
| --- | --- | --- | --- | --- | --- | --- | --- | --- | --- | --- | --- | --- | --- |
|  | AVE | Composite Reliability | Cronbach Alpha | PUS | PEU | CSE | SI | AIT | EN | RF | CF | AT | BI |
| PUS | 0.70 | 0.90 | 0.85 | **0.84** | | | | | | | | | |
| PEU | 0.69 | 0.87 | 0.78 | 0.43 | **0.83** | | | | | | | | |
| CSE | 0.63 | 0.87 | 0.80 | 0.30 | 0.51 | **0.79** | | | | | | | |
| SI | 0.67 | 0.80 | 0.51 | 0.40 | 0.41 | 0.36 | **0.82** | | | | | | |
| AIT | 0.54 | 0.78 | 0.59 | 0.27 | 0.51 | 0.40 | 0.29 | **0.73** | | | | | |
| EN | 0.75 | 0.90 | 0.83 | 0.66 | 0.42 | 0.39 | 0.39 | 0.29 | **0.87** | | | | |
| RF | 0.68 | 0.86 | 0.76 | 0.39 | 0.31 | 0.25 | 0.33 | 0.38 | 0.53 | **0.82** | | | |
| CF | 0.60 | 0.82 | 0.66 | 0.58 | 0.35 | 0.38 | 0.42 | 0.28 | 0.64 | 0.56 | **0.77** | | |
| AT | 0.79 | 0.92 | 0.86 | 0.67 | 0.41 | 0.41 | 0.37 | 0.20 | 0.80 | 0.53 | 0.63 | **0.89** | |
| BI | 0.81 | 0.93 | 0.89 | 0.57 | 0.34 | 0.30 | 0.43 | 0.15 | 0.67 | 0.43 | 0.46 | 0.66 | **0.90** |

## 5.2 Structural Model

In order to test the statistical significance of the relationships in the model a bootstrap procedure with 200 re-samples was used. In PLS-SEM, we try to explain as much of the variance in any construct by the variation in antecedent constructs. The results for the model as shown in Fig. 1 are summarized in Fig. 4. There we give the path coefficient (i.e., the strength of the coupling in the SEM model), as well as the *t*-test values. The relevant critical values for a two-tailed test are: *t*=1.96 has *p*=0.05, and *t*=2.58 has *p*=0.01. Here we use *t*-values higher than 1.96 as significant, and reject paths with lower probability than 0.05 and corresponding lower *t* vaLues.

# 6 Discussion

We find that the antecedent constructs explain 53% of the variance in 'behaviour intention'. At an intermediate stage of the model, 68% of the variance in attitude, 49% of the variance in 'enjoyment', 40% of the variance in 'perceived usefulness' and 37% of the variance in 'perceived ease of use' are explained in a similar way. More importantly, we see that 'attitude' and 'enjoyment' have an important effect on predicting usage intentions. This shows that, even for detailed mathematical assessments, it is important for students to take pleasure in and appreciate the on-line assessments. It also shows that their opinions and feelings about the assessments make a difference to their usage intentions. On the other hand, we noticed that the construct 'perceived usefulness' has only an indirect effect through the 'attitude'. . This conclusion matches with the acceptance model for computer based assessment (CBAAM) of Terzis & Economides (2011) who also find that perceived usefulness has no direct effect on behavioural intention to use a computer-based assessment.

'Attitude' in turn is predicted by 'perceived usefulness' and 'enjoyment', but 'enjoyment' has a stronger effect than 'perceived usefulness'. This could indicate that an on-line test has to include fun activities, challenging activities that make students find them enjoyable while they are learning mathematics. We see that creating teaching strategies that include enjoyable activities is essential to enhance student's learning in mathematics. Making good use of games, quizzes, and other creative approaches to create more enjoyment and interest in learning mathematics is essential.

On the other hand, 'comparative feedback' has a strong direct effect on predicting 'perceived usefulness' and 'enjoyment'. This confirms that students value the fact of getting on-line feedback, since they find that it is both useful and enjoyable. This is also demonstrated by looking at the relevance of the indirect effects by evaluating their *t* values. The effect of 'comparative feedback' on attitude and on usage intentions are both highly relevant, as can be seen in 'Received feedback' has an important effect on 'enjoyment' but not on 'perceived usefulness', as can be seen by the direct *t*-values, $t_{RF \to EN}$ = 1.99 and $t_{RF \to PUS}$ = 0.56. This seems to indicate that students enjoy receiving on-line feedback in mathematical assessement. The indirect effects on attitude ($t_{RF \to AT}$ = 1.85) and usage intentions ($t_{RF \to BI}$ = 1.64) demonstrate that this factor does not crucial to trigger a favourable behavioural intention, see Table 4.

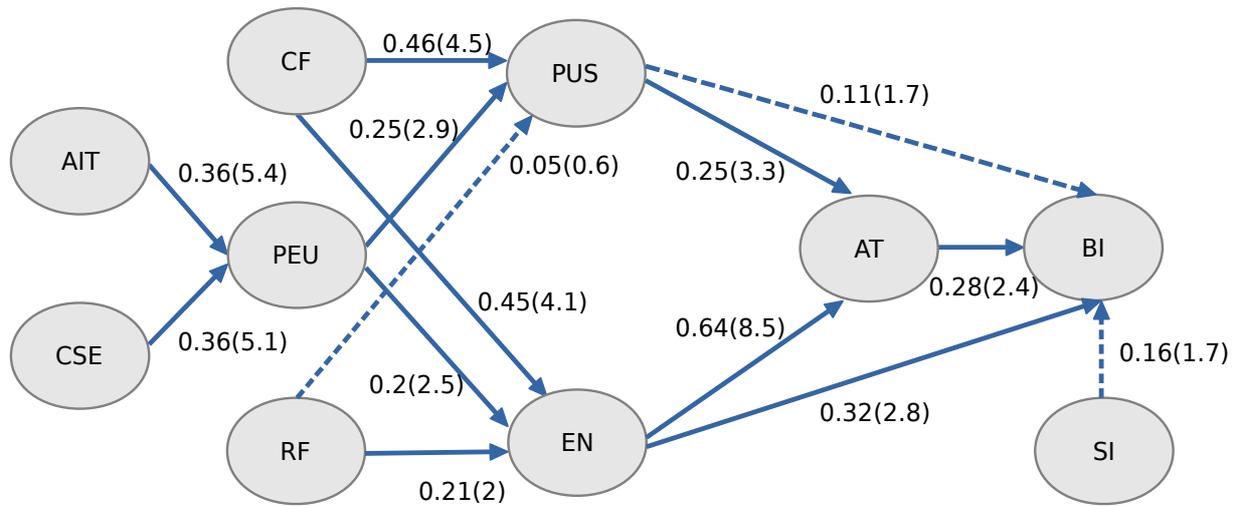

*Figure 4. A visual representation of the results obtained from the PLS-SEM model for student responses. The arrows represent hypothetical relationships between the latent variables; if solid they are accepted; dashed ones are rejected. The number next to each arrow shows the path coefficient, with the t-value in parentheses.*

Table 4. Respect to enjoyment this finding matches with the study (Moon & Kim, 2001) who find that enjoyment has a positive impact on behavioural intentions.

Table 4: *t*-values for indirect effects.

| Independent construct | Dependent construct | | | |
|---|---|---|---|---|
| | PUS | EN | AT | BI |
| AIT | 2.55 | 2.30 | 2.66 | 2.6 |
| CSE | 2.44 | 2.06 | 2.37 | 2.32 |
| PEU | | | 3 | 2.9 |
| CF | | | 4.75 | 4.07 |
| RF | | | 1.85 | 1.64 |

Perceived ease of use has a strong influence on perceived usefulness ($t_{PEU \to PUS}$ = 2.93) and enjoyment ($t_{PEU \to EN}$ = 2.52). This also contributes to the strong indirect effect on attitude ($t_{PEU \to AT}$ = 3) and usage intentions ($t_{PEU \to BI}$ = 2.9). These match what we mentioned earlier: perceived enjoyment and perceived usefulness mediate the effects of perceived ease of use on intention. This means that this factor is essential to obtain positive attitude and usage intentions. This is proven by following the causal chain (PEU → EN → AT → BI) that could indicate that if students perceive it as easy to use, they are more likely to have an enjoyable experience, and they are more willing to use it. The causal links (PEU → PUS → AT → BI) shows that when technology is easier to operate, it is more useful, and therefore, students are more willing to apply it. Enjoyment and perceived usefulness mediate the impact of three constructs (perceived feedback, comparative feedback, perceived ease of use) on attitude this has a direct effect on usage behaviour. This reveals that enjoyment and perceived ease of use are powerful factors for predicting usage intentions. This is consistent with what Davis *et al.* (1992) point out "enjoyment will explain significant variance in usage intentions beyond that accounted for by usefulness alone." On the other hand, perceived ease of use has been hypothesized as an important factor influencing usage behaviour (Davis, 1989).

Availability of information technology and computer self-efficacy have a strong effect on perceived ease of use. This could indicate that students who encounter some difficulties during an on-line test (regarding system's operation or questions' content) need technical support such as help-desks, on-line support services, guidance by the IT staff and faculty to overcome these situations. Therefore, when technical assistance is provided, it is more likely to find using the on-line environment easier. This is supported by Terzis & Economides (2011) who explain that in the context of computer-based assessment, the availability of information technology determines perceived ease of use. We also see that students who feel comfortable using computers, will find it easier for mathematical on-line assessments.

These results show that the indirect relationships AIT → AT, AIT → BI, AIT → EN, AIT → PUS , CF → AT, CF →BI , CSE → AT, CSE → BI, CSE → EN, CSE → PUS, PEU → AT, PEU → BI are all significant for the model. Furthermore, using the measurement model, especially the significance of indicator variables, we see that items AT1, AT2, BI2, BI3, CF3, EN3, and PUS1 are the most important in the model. If we quickly look at what these specific questions ask, we see that AT1 "I like doing on-line test and exams in subjects that require mathematical answers" and AT2 "I look forward to those aspects of my learning of mathematics that require me to use on-line assessment" suggest that we need to make sure that the current tests satisfies the students expectations as well as possible.  The questions CF3 "on-line feedback helps me better understand mathematical subjects", EN3 "doing mathematical on-line tests is enjoyable" and PUS1 "I find on-line tests useful to support my learning of mathematical subjects"  are also statements about the quality of the current tests, and clear;y show those aspects test development should focus on—enjoyment, usefulness, and quality feedback. The importance of  items BI2 "I would like to continue my use of on-line assessment to support my learning of mathematics" and BI3 "all things considered, I expect to continue doing on-line test or exams to assist my learning of mathematics" are more troublesome—we do know it is a problem, and as yet largely unsolved, to develop online assessments for material beyond that used in the first year of the English university curriculum.

# 7 Conclusions

Immediate on-line feedback is an essential feature of the formative assessment, since it has positive effects on students' learning performance. It activates their intrinsic motivation (Dreher, Reiners, & Dreher, 2011) and therefore helps them achieve their goals (Whitelock & Brasher, 2006). In this study we have shown the important effects, both direct and indirect, of on-line feedback on students' attitude and usage intentions in advanced mathematical assessments.

We have found that 'comparative feedback' has a strong effect on predicting 'perceived usefulness'  and 'enjoyment'. This demonstrates the value students attach  to on-line feedback, and demonstrates the most important message from this study: we need to combine utility with enjoyment in a successful assessment, even in "dry" subject such as application of advanced mathematics. This is also demonstrated by looking at the strong positive indirect effects of 'comparative feedback' on 'attitude 'and 'usage intentions'.

We note that 'received feedback' has a strong influence on 'enjoyment' but not on 'perceived usefulness'.  This indicates the rather surprising result that students find the experience of receiving feedback in an on-line (mathematical) test more enjoyable than useful. Also, the indirect effects on 'attitude' and 'behavioural intention' have a small *t* value, and as such the process of receiving feedback is less important to the attitude and acceptance of on-line testing than we expected.

Nevertheless, our results also reveal that 'attitude' and 'enjoyment' are the most important factors influencing 'usage intention'. To our surprise 'usefulness' does not have a direct effect—in a future study we hope to report on instructors attitudes, but indications are that usefulness for students is an important driver for instructors. On the other hand, 'usefulness' has an indirect effect: together with enjoyment it strongly predicts 'attitude'.

In its turn, 'perceived ease of use' is an important factor influencing 'usefulness' and 'enjoyment'. This also contributes to its strong indirect effect on 'attitude' and 'usage intentions'. Therefore, 'enjoyment' and 'perceived ease of use' are the most powerful factors for predicting usage intentions. We know that 'availability of information technology' and 'computer self-efficacy' have a strong effect on perceived ease of use. Taking into account these factors can be the best way to design a maths e-assessment activity for UK students.

We would therefore recommend to start by giving adequate training to the students before using the assessment engines, so that they can make most effective use of the assessment. One needs to check carefully under which circumstances, both hardware and software based, such assessments have problems, and steer students to a sufficient number of IT resources that deal with these assessments well. Assessments need to be well designed, designing tasks that students find more enjoyable than useful. It would probably be a good idea to study what tasks students do find enjoyable, and which they dislike. Feedback needs to be good, but is not as important as we though at the start of this study.

# Appendix A

The questions put to the students are set out in Table 5, where we list each question in detail.

Table 5: A list of all the questions in the questionnaire answered by students, grouped under there respective categories

| |
|---|
| **Perceived Usefulness (PUS)** |
| PUS1. I find on-line tests useful to support my learning of mathematical subjects |
| PUS2. Doing on-line tests enhance my mathematical knowledge |
| PUS3. On-line assessments help me to understand mathematical topics better |
| PUS4. I find it useful that I can answer mathematical on-line test at any time and at a place of my choice |
| **Perceived Ease of Use (PEU)** |
| PEU1. My interaction with the systems providing mathematical on-line assessments is clear and understandable |
| PEU2. I find it straightforward to use on-line assessments to support my learning of mathematics |
| PEU3. It is straightforward to become skilful at using on-line assessments of mathematical subjects |
| **Computer Self-Efficacy (CSE)** |
| CSE1. I feel confident doing mathematical on-line assessment |
| CSE2. I feel comfortable using mathematical on-line assessments on my own |
| CSE3. I am able to use on-line assessments of mathematics even if there is no one around to explain me how to use the system |
| CSE4. In general, I feel confident doing on-line assessment |
| **Social Influence (SI)** |
| SI1. My lecturer expects me to do on-line tests that contain mathematics |
| SI2. Classmates are positive about the use of mathematical on-line assessments |
| **Availability IT Services (AIT)** |
| AIT1. When I need help to learn to use on-line assessment specialised university staff is there to support me |
| AIT2. Internet speed at my university is fast enough to use an on-line learning environment |
| AIT3. My university has enough computing infrastructure to support on-line testing |
| **Enjoyment (EN)** |
| EN1. I enjoy using on-line assessments that require mathematical answers |
| EN2. Using on-line assessment of mathematics stimulates my curiosity |
| EN3. Doing mathematical on-line tests is enjoyable |
| **Received Feedback (RF)** |
| RF1. The on-line feedback returned with my mathematical exercises and exams were fair & balanced |
| RF2. On-line feedback gave me enough information on where I went wrong in mathematical exercises and exams |
| RF3. From my on-line feedback, I learnt how to improve my work for mathematical subjects |
| **Comparative Feedback (CF)** |
| CF1. On-line feedback helps me resolve faster doubts about the mathematical material than traditional feedback |
| CF2. Electronic assessments of my mathematical subjects allow me to get grades faster, so I know if I am doing well in my topic |
| CF3. On-line feedback helps me better understand mathematical subjects |
| **Attitude (AT)** |
| AT1. I like doing on-line test and exams in subjects that require mathematical answers |
| AT2. I look forward to those aspects of my learning of mathematics that require me to use on-line assessment |
| AT3. On-line tests that contain mathematics are useful |
| **Behavioural Intention (BI)** |
| BI1. I will use electronic tools to support my learning of mathematical subjects in the future |
| BI2. I would like to continue my use of on-line assessment to support my learning of mathematics |
| BI3. All things considered, I expect to continue doing on-line test or exams to assists my learning of mathematics |

# Appendix B

As an example of the capabilities we expose our students, consider the simplified problem discussed in Fig. 5. There we show a highly simplified example of the type of feedback given to students, some of which is based on known common mistakes (malrules), such as differentiating when asked to integrate, mistakes in carrying through factors in functions (i.e., functions of functions), etc. In masteringphysics our approach is much the same; in addition we also use analysis based on the presence of terms containing the relevant physical parameters.

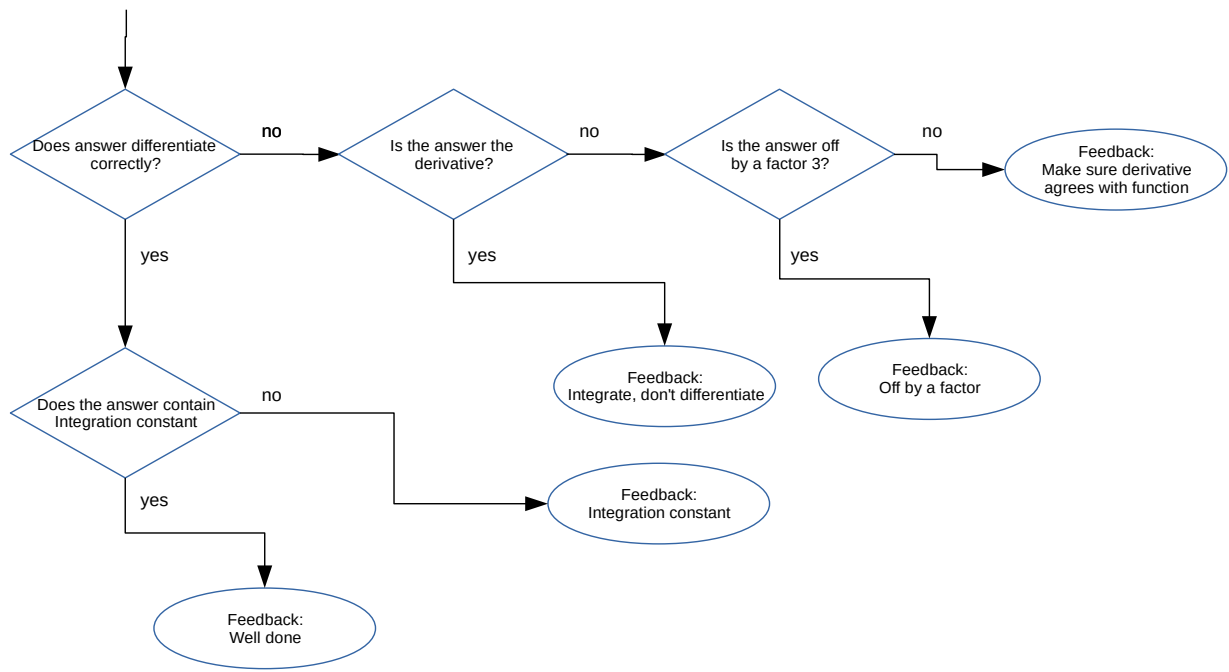

*Figure 5. flow diagram representation of a simplified stack question. We assume that the student is asked to perform the indefinite integrate the function cos(6 x), and have shown some of the feedback that can be given. Note that this is actually an extremely simple example of the type of feedback a student may be exposed to.*